\documentclass[sigconf]{acmart}

\usepackage{graphicx}
\usepackage{subcaption}

\AtBeginDocument{%
  \providecommand\BibTeX{{%
    \normalfont B\kern-0.5em{\scshape i\kern-0.25em b}\kern-0.8em\TeX}}}

\copyrightyear{2026}
\acmYear{2026}
\setcopyright{cc}
\setcctype{by}
\acmConference[ETRA '26]{2026 Symposium on Eye Tracking Research and Applications}{June 01--04, 2026}{Marrakesh, Morocco}
\acmBooktitle{2026 Symposium on Eye Tracking Research and Applications (ETRA '26), June 01--04, 2026, Marrakesh, Morocco}
\acmDOI{10.1145/3797246.3805854}
\acmISBN{979-8-4007-2519-7/2026/06}

\begin{document}

\title{Exploratory Integration of EEG Spectral Features and Gaze Variability for Mild Cognitive Impairment Discrimination}


\author{Takeru Mukunoki}
\orcid{0009-0003-0294-1728}
\email{2177145z@gsuite.kobe-u.ac.jp}
\affiliation{%
  \institution{Kobe University}
  \city{Kobe}
  \country{Japan}
}

\author{Mamoru Hiroe}
\orcid{0000-0002-5145-5406}
\email{hiroe@ieee.org}
\affiliation{%
  \institution{Osaka Seikei University}
  \city{Osaka}
  \country{Japan}
}
\affiliation{%
  \institution{Kobe University}
  \city{Kobe}
  \country{Japan}
}

\author{Minoru Nakayama}
\orcid{0000-0001-5563-6901}
\email{nakayama@meiji.ac.jp}
\affiliation{%
  \institution{Meiji University}
  \city{Tokyo}
  \country{Japan}
}

\author{Yujia Zheng}
\orcid{0000-0003-0780-9462}
\email{218k215k@stu.kobe-u.ac.jp}
\affiliation{%
  \institution{Kobe University}
  \city{Kobe}
  \country{Japan}
}

\author{Yuma Sonoda}
\orcid{0000-0001-8441-7441}
\email{yuma@dragon.kobe-u.ac.jp}
\affiliation{%
  \institution{Kobe University}
  \city{Kobe}
  \country{Japan}
}

\author{Hisatomo Kowa}
\orcid{0000-0002-5300-8589}
\email{kowa@med.kobe-u.ac.jp}
\affiliation{%
  \institution{Kobe University}
  \city{Kobe}
  \country{Japan}
}

\author{Takashi Nagamatsu}
\orcid{0000-0001-8426-5978}
\email{nagamatu@kobe-u.ac.jp}
\affiliation{%
  \institution{Kobe University}
  \city{Kobe}
  \country{Japan}
}

\renewcommand{\shortauthors}{Mukunoki and Hiroe, et al.}

\begin{abstract}

Early detection of mild cognitive impairment (MCI) is an important challenge in aging societies.
Electroencephalography (EEG) and eye-tracking have independently been explored as potential biomarkers; however, their integrative effects remain insufficiently examined.
This exploratory study investigated whether combining EEG spectral features with gaze variability may provide complementary information for MCI discrimination.
EEG signals were recorded using the 10--20 system, and spectral power features were extracted.
We compared three models:
(a) high-dimensional EEG features,
(b) L1-regularized feature selection (LASSO),
and (c) integration of the selected EEG features with gaze variability.
Performance was evaluated using leave-one-out cross-validation and area under the ROC curve (AUC).
Model (a) yielded limited discrimination (AUC = 0.52).
Feature selection increased AUC (0.64), and additional integration of gaze variability further increased AUC (0.78).
These preliminary findings suggest potential complementarity between neural and behavioral variability measures.
\end{abstract}





\maketitle

\section{Introduction}

Mild cognitive impairment (MCI) is regarded as an intermediate stage between normal aging and dementia, and its early detection is clinically and socially important. 
Conventional cognitive assessments widely used today often require face-to-face administration and evaluation by trained professionals. 
Therefore, the development of simple and objective digital indicators is increasingly required.

Electroencephalography (EEG) has been widely investigated as a technique for capturing neural activity alterations associated with dementia and MCI. 
In contrast, eye tracking has also been suggested as a potential indicator of cognitive decline, particularly through measures such as saccadic characteristics and fixation stability.

However, studies that integrate both EEG and gaze measures and quantitatively evaluate changes in discrimination performance remain limited. 
In this study, we investigate whether integrating EEG spectral features with gaze variability during a fixation task can improve discrimination between individuals with MCI and healthy controls.

\section{Related work}

EEG has been widely investigated as a non-invasive biomarker for the early detection of MCI and Alzheimer's disease. 
In particular, studies focusing on spectral band power in resting-state EEG have reported band-specific alterations, such as increased theta power and decreased beta power \cite{Cassani2018,Meghdadi2024}. 

Eye movements have also attracted attention as behavioral indicators for MCI detection. 
Measures such as saccadic characteristics and fixation stability have been reported to relate to cognitive decline \cite{Wolf2023,Shah2025}. 
However, most studies have examined gaze-based indicators alone, and integrative evaluations with neural activity remain limited.

Simultaneous recording of EEG and eye movements has been explored for analyzing the relationship between neural activity and gaze behavior \cite{Kastrati2021}. 
However, studies that examine their combined effects for clinical discrimination, such as MCI classification, remain limited.

\section{Method}

\subsection{Participants}

This study included 38 participants (17 MCI, 21 healthy controls). 
Group classification was based on the Montreal Cognitive Assessment (MoCA) score.
The mean age of the MCI group was 75.9 $\pm$ 3.1 years and that of the healthy control group was 74.8 $\pm$ 6.0 years.

\subsection{Task}

Participants performed a fixation task in which they were instructed to fixate on a single point presented on a screen. 
Each trial lasted 20 seconds, and the first 4 seconds of the recorded data were used for analysis.

\subsection{Data Acquisition and Feature Extraction}

\subsubsection{EEG Features}
EEG signals were recorded using an EEG-1214 system (Nihon Kohden Corp.) according to the international 10–20 system with the earlobes used as the reference. 
For analysis, signals were re-referenced by subtracting regional reference electrodes (Fz for frontal, C3/C4 for central, and Pz for posterior electrodes).
The sampling rate was 500 Hz.
For each participant, time-series signals were acquired for 20 seconds, and the power spectral density was estimated using Welch's method.

For all channels, the average power of multiple frequency bands (delta, theta, alpha, low-beta, high-beta, and gamma) was calculated and used as the high-dimensional EEG feature set in model (a).
Subsequently, L1 regularization (LASSO) was applied as an exploratory analysis using the full dataset to identify the two most contributive features. These selected features were used in models (b) and (c).
Because this feature selection was performed exploratorily, the stability of the selected electrodes and frequency bands should be verified in future studies.

\subsubsection{Gaze Features}
Gaze data were recorded using an EyeLink Portable Duo (SR Research Ltd.) at a sampling rate of 1000 Hz.
Gaze coordinate data recorded during the fixation task were analyzed after excluding saccade intervals. 
The standard deviations of gaze position in the $x$ and $y$ directions were calculated as indicators of fixation stability.

\subsection{Classification Analysis}

Three conditions were compared in a stepwise manner:

\begin{itemize}
\item (a) A model using all EEG features
\item (b) A model using the two EEG features selected by L1 regularization (LASSO)
\item (c) A model using the selected EEG features together with gaze features (R\_std\_x, L\_std\_x)
\end{itemize}

Classification was performed using logistic regression. 
To address class imbalance between the MCI ($n = 17$) and healthy control ($n = 21$) groups, the class weights were balanced.

Performance was evaluated using leave-one-out cross-validation (LOOCV). 
In each fold, one participant was used as the test sample and the remaining participants were used for model training. 
The predicted probabilities from all folds were aggregated to compute the receiver operating characteristic (ROC) curve and the area under the curve (AUC).

Feature selection using L1 regularization (LASSO) was conducted exploratorily using the full dataset, and the selected features were used in models (b) and (c).
Because feature selection and performance evaluation were not fully separated, the reported AUC values should be interpreted as preliminary upper-bound estimates.

\section{Results}
\begin{figure*}[htbp]
\centering

\begin{subfigure}{0.32\linewidth}
    \centering
    \includegraphics[width=\linewidth]{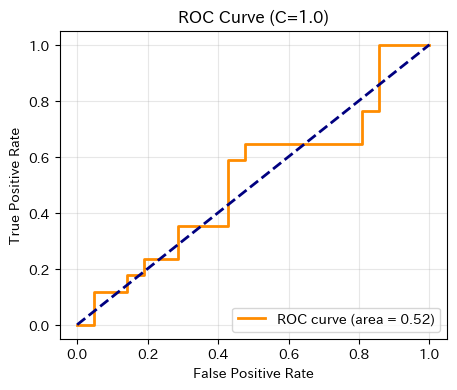}
    \caption{All EEG features \\ AUC = 0.52}
\end{subfigure}
\hfill
\begin{subfigure}{0.32\linewidth}
    \centering
    \includegraphics[width=\linewidth]{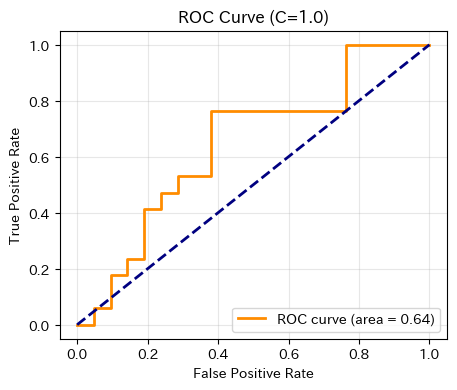}
    \caption{LASSO-selected EEG features \\ AUC = 0.64}
\end{subfigure}
\hfill
\begin{subfigure}{0.32\linewidth}
    \centering
    \includegraphics[width=\linewidth]{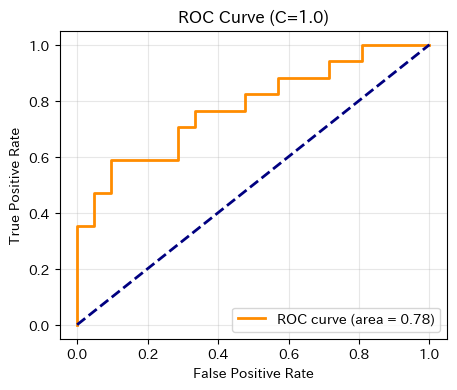}
    \caption{LASSO-selected EEG + Gaze features \\ AUC = 0.78}
\end{subfigure}

\caption{
ROC curves for the three classification models. 
Performance improved stepwise from (a) to (c).
}
\label{fig:roc_comparison}

\end{figure*}

Figure~\ref{fig:roc_comparison} shows the ROC curves for the three models.
Model (a) yielded limited discrimination (AUC = 0.52).
Model (b) improved performance (AUC = 0.64).
The selected features corresponded to the delta-band power at electrode O1 and the low-beta-band power at electrode T6.
Model (c) further increased the AUC to 0.78.

These results suggest that dimensionality reduction may improve classification compared with the full EEG feature model. Furthermore, integrating gaze features resulted in an additional improvement (A gaze-only model yielded an AUC of 0.55.).

In particular, the integration of gaze features increased the AUC from 0.64 (LASSO-selected EEG) to 0.78, suggesting a substantial improvement in classification performance.

\section{Discussion}

In this study, the integration of EEG variability features and gaze variability during fixation suggested the potential to improve discrimination between individuals with MCI and healthy controls.

The occipital cortex is involved in visual processing, and fixation tasks require both visual information processing and attentional control. 
Variability in neural activity and instability of gaze may therefore reflect different aspects of a common cognitive process.

This study has several limitations. 
The sample size was limited, and external validation was not conducted. 
Future studies should include larger participant cohorts and evaluate the generalizability of the proposed approach across different tasks. 
The current study focused on minimal gaze features to maintain a simple experimental setup, while richer gaze descriptors (e.g., fixation dispersion or microsaccade-related features) should be explored in future work.
In addition, further investigation of spectral features and other multimodal indicators may help clarify the mechanisms underlying the observed improvements in classification performance.

\section{Conclusion}

This study explored whether integrating EEG spectral features and gaze variability could improve discrimination between individuals with mild cognitive impairment (MCI) and healthy controls. 
The results showed that classification performance improved when dimensionality reduction was applied to EEG features and further improved when gaze features were incorporated. 
These findings suggest that combining neural and behavioral indicators may provide complementary information for detecting subtle cognitive changes associated with MCI. 
Future work should validate the proposed approach with larger datasets and investigate additional multimodal features.


\balance
\begin{acks}
This work was supported by JSPS KAKENHI Grant Number 22K18423.
\end{acks}


\bibliographystyle{ACM-Reference-Format}
\bibliography{references}

\end{document}